\documentclass[onecolumn]{aastex631}

\newcommand{\target}{KSP-OT-201701a}

\newcommand{\hi}{\hbox{H\,{\sc i}}}

\newcommand{\ha}{\hbox{H$\alpha$}}
\newcommand{\hei}{\hbox{He\,{\sc i}}}

\begin{document}

\title{Discovery of a Short-Period and Unusually Helium-Deficient Dwarf Nova \target\ by the KMTNet Supernova Program}
\shorttitle{Lee et al.}

\correspondingauthor{Youngdae Lee}
\email{hippo206@gmail.com}

\author[0000-0002-6261-1531]{Youngdae Lee} 
\affil{Department of Astronomy and Space Science, Chungnam National University, Daejeon 34134, Republic of Korea}
\affil{Korea Astronomy and Space Science Institute, 776, Daedeokdae-ro, Yuseong-gu, Daejeon 34055, Republic of Korea}

\author[0000-0001-9670-1546]{Sang Chul Kim} 
\affil{Korea Astronomy and Space Science Institute, 776, Daedeokdae-ro, Yuseong-gu, Daejeon 34055, Republic of Korea}
\affil{Korea University of Science and Technology (UST), Daejeon 34113, Republic of Korea}

\author[0000-0003-4200-5064]{Dae-Sik Moon}
\affil{David A. Dunlap Department of Astronomy and Astrophysics, University of Toronto, 50 St. George Street, Toronto, ON M5S 3H4, Canada}

\author[0000-0002-3505-3036]{Hong Soo Park}
\affil{Korea Astronomy and Space Science Institute, 776, Daedeokdae-ro, Yuseong-gu, Daejeon 34055, Republic of Korea}
\affil{Korea University of Science and Technology (UST), Daejeon 34113, Republic of Korea}

\author[0000-0001-7081-0082]{Maria R. Drout}
\affil{David A. Dunlap Department of Astronomy and Astrophysics, University of Toronto, 50 St. George Street, Toronto, ON M5S 3H4, Canada}

\author[0000-0003-3656-5268]{Yuan Qi Ni}
\affil{David A. Dunlap Department of Astronomy and Astrophysics, University of Toronto, 50 St. George Street, Toronto, ON M5S 3H4, Canada}

\author[0000-0002-5742-8476]{Hyobin Im}
\affil{Korea Astronomy and Space Science Institute, 776, Daedeokdae-ro, Yuseong-gu, Daejeon 34055, Republic of Korea}
\affil{Korea University of Science and Technology (UST), Daejeon 34113, Republic of Korea}

\shorttitle{Dwarf nova : {\target}}
\shortauthors{Lee et al.}

\begin{abstract}
We present the first ever discovery of a short-period and unusually helium-deficient dwarf nova 
\target\ by the Korea Microlensing Telescope Network Supernova Program.
The source shows three superoutbursts, each led by a precursor outburst,
and several normal outbursts in $BVI$ during the span of $\sim$ 2.6 years
with supercycle and normal cycle lengths of about 360 and 76 days, respectively.
Spectroscopic observations near the end of a superoutburst 
reveal the presence of strong double-peaked {\hi} emission lines 
together with weak {\hei} emission lines. 
The helium-to-hydrogen intensity ratios measured by 
${\hei}_{\lambda5876}$ and ${\ha}$ lines are 0.10 $\pm$ 0.01 at a quiescent phase and 0.26 $\pm$ 0.04 at an outburst phase,
similar to the ratios found in long-period dwarf novae while
significantly lower than those in helium cataclysmic variables (He CVs).
Its orbital period of 51.91 $\pm$ $2.50$ minutes, which is estimated based on time series spectroscopy,
is a bit shorter than the superhump period of 56.52 $\pm$ $0.19$ minutes, 
as expected from the gravitational interaction between the eccentric disk 
and the secondary star.
We measure its mass ratio to be $0.37^{+0.32}_{-0.21}$ using the superhump period excess of $0.089\pm0.053$.
The short orbital period, which is under the period minimum,
the unusual helium deficiency, and the large mass ratio
suggest that \target\ is a transition object evolving to a He CV
from a long-period dwarf novae with an evolved secondary star.
\end{abstract}


\section{Introduction} \label{sec:intro}

Dwarf novae are binary systems composed of a white dwarf (WD) 
as the primary and a low-mass star as the secondary \citep{War95},
accompanied by the accretion disk formed by mass-flow from the secondary.
The increased temperature of the accretion disk resulting from 
mass accretion triggers outbursts that appear as dwarf novae \citep{Can86,Sma84,Osa96}.  
The evolution of dwarf novae is mainly determined by 
the loss of angular momentum driven by the gravitational radiation from their
binary orbital motions \citep{Pac81,Rap82}.
The orbital period and binary separation distance decrease as they evolve
until they reach the so-called period minimum of about 80 minutes, 
at which point the gravitational radiation time scale becomes similar to the Kelvin-Helmholtz time scale.
After the period minimum,
the radius of the secondary star as well as the binary separation distance and orbital period 
start to increase, known as the periodic bouncer \citep{Rap82}.

Of particular interests among dwarf novae is the origin and evolution of the small number 
of dwarf novae showing orbital periods below the period minimum.
They occupy $\sim$ 5\% of the entire population of dwarf novae \citep{Pal20},
and are thought to have short periods because the increased density of the secondary 
resulting from the loss of an extended atmosphere shrinks the binary orbital motion \citep{Pod03}.
This process is more effective for an evolved secondary due 
to its already increased density near the center.
The short-period dwarf novae are categorized into two subclassess:
AM Canum Venaticorum (AM CVn) systems and He-rich  cataclysmic variables (He CVs).
The former show strong He emission with no H emission \citep{And05}, whereas
the latter do both He and H emission together with a larger 
${\hei}_{\lambda5876}/{\ha}$ flux ratio \citep[He/H; ][]{Ken15}.
In comparison, dwarf novae above the period minimum show the flux ratio 
in the range of 0.15--0.30 \citep{Wil82,Rit03,Tho04}.
The majority of the short-period dwarf novae are AM CVn systems,
with He CVs being extremely rare and accounting for only $\sim$ 7~\% of them \citep{Bre12}. 

Three different binary types have been suggested for the short-period dwarf novae:
(1) binaries of two WDs;
(2) binaries of a WD and a He-burning star with a stripped H envelope, and
(3) binaries of a WD and an evolved star off the main sequence \citep[see][and references therein]{Gre20}.
The first two types, which pass through two common envelope phases during their evolution, 
are known to be the origin of AM CVn showing no H emission. 
The third type, on the other hand, is known to be the origin of AM CVn and all of the He CVs 
\citep{Pod03,Gol15,Gre20,ElB21}. According to \citet{Ken15}, 
a long-period binary composed of a WD and an evolved star off the main sequence
very likely passes through a phase of He CV on its eventual evolution to an AM CVn
by gradually losing H and showing higher He/H ratios as they evolve.

One of the key observations that can potentially confirm this evolution scenario of dwarf novae
is the identification of an intermediate example that shows a short period
but with a smaller He/H ratio than what has been observed in He CVs. 
To the best of our knowledge, no short-period dwarf nova 
below the period minimum has been observed with a He/H ratio $<$ 0.5.
In this {\it Letter}, we present the first discovery of 
which we name \target.
We suggest that this object is most likely evolving from a long-period binary 
consisting of a WD and an evolved star to the short-period He CV phase.

\section{Discovery of \target\ and Light Curves} \label{sec:discovery}

\target\ was discovered by the KMTNet (Korea Microlensing Telescope Network) Supernova Program 
\citep[KSP;][]{Moo16} during its wide-field (= 2\degr\ $\times$ 2\degr\ at a pixel sample rate of 0\farcs4 per pixel) 
monitoring observations of the spiral galaxy 
NGC 2280 for about 2.6 years between 2016 October and 2019 May.
KSP conducts high-cadence, multi-color (= $BVI$ bands) observations  
optimized for discovering and monitoring early supernovae \citep[e.g., ][]{Afs19,Moo21}, taking advantage of 24-hour continuous sky coverage provided by the three 1.6-m telescopes
of the KMTNet \citep{Kim16} in Chile, South Africa and Australia.
Such observations are also useful 
for understanding the origins and population of 
classical novae and dwarf novae \citep{Ant17,Bro18,Lee19}.

\begin{figure*}
\epsscale{1.2}
\plotone{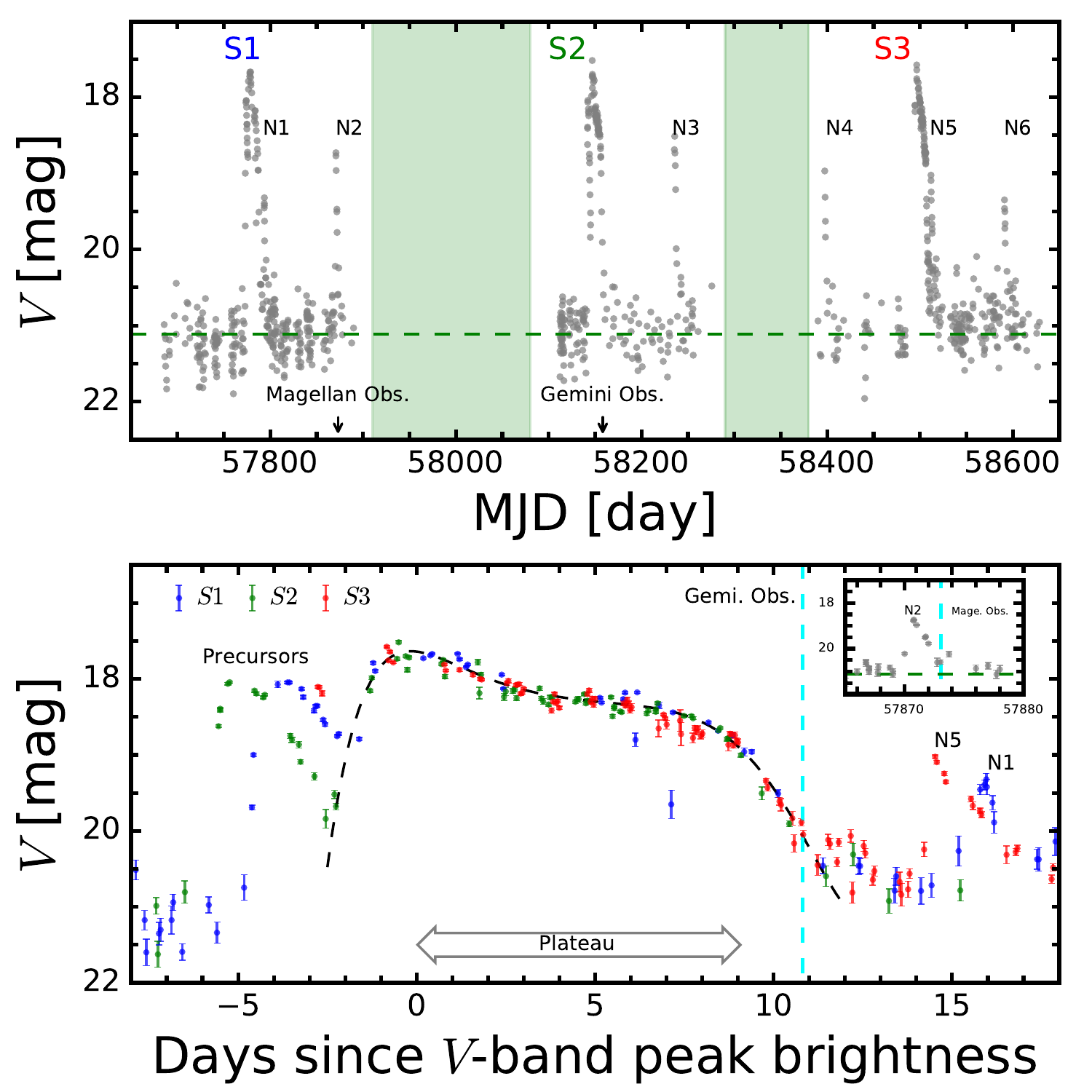}
\caption{({\it Top Panel}) The observed $V$-band light curve of \target\ for $\sim$2.6 years
showing three superoutbursts (S1--S3) as well as six normal outbursts (N1--N6).
The horizontal dashed line represents the mean quiescent magnitude $V$ = 21.11 mag.
Two arrows indicate the epoch of spectroscopic observations.
Green shades represent the ephochs when the source was near the Sun.
({\it Bottom Panel}) Three superoutbursts S1 (blue dots), S2 (green dots), and S3 (red dots)
are compared with the best-fit polynomial (black dashed line).
Their plateau duration is indicated by the double arrow.
({\it Inset Panel}) The observed $V$-band light curve for $N2$ is shown.
The vertical dashed lines indicate the epoch of spectroscopic observations.
}
\label{LCs}
\end{figure*}

Figure~\ref{LCs} (top panel) shows the entire $V$-band light curve of \target,
which is located at $(\alpha, \delta)_{\rm J2000} = 
(06^h\,39^m\,23.2^s, -26^\circ\,37'\,18.8'')$\footnote{$(l,b)=(235.84662^{\circ},-14.24921^{\circ})$},
obtained with the KMTNet. 
The integration time of each exposures is 60 seconds, and the mean cadence is 
about 8 hours for each of $BVI$ bands.
We use DAOPHOT-based point-spread function (PSF) photometry \citep[DAOPHOT II;][]{Ste87} 
to obtain the fluxes of the source that were calibrated against
nearby $BVi'$ standard stars from the AAVSO Photometric All-Sky Survey database (APASS).
We adopt the relation $I = i' - 0.4$ to convert the 
APASS $i'$-band standard photometric system to the KMTNet $I$-band system \citep[see][for the details]{Par17}.
We apply the extinction correction of $A_B = 0.51$ mag, $A_V = 0.40$ mag, and $A_I = 0.25$ mag
using the extinction $E(B-V) = 0.13$ mag toward {\target}
in the Galactic extinction model of \citet{Sch98} with the
extinction reddening law of \citet{Car89} with $R_V = 3.1$.
In this paper, we use extinction-corrected magnitudes and colors unless otherwise specified.

In the top panel of Figure~\ref{LCs}, we can identify three 
apparent superoutbursts (which we call $S1$, $S2$, and $S3$) 
featured with the presence of a plateau of $\sim$9 days 
as shown in the bottom panel as well as six normal outbursts ($N1$--$N6$).
We determine the mean quiescent magnitude of \target\ to be 
$21.22\pm0.20$, $21.11\pm0.20$, and $20.58\pm0.21$ mag in the $B$, $V$, and $I$ bands 
using the images obtained when there was no outburst activity.
This gives a mean supercycle ($P_{sc}$) period of 360 days for \target. 
However, considering the gap of the data when the source was near the Sun (green shades in the top panel of Figure~\ref{LCs}), 
we note that we cannot rule out the possibility of $P_{sc}$ = 180 days.
The mean cycle of normal outbursts is 78.5 days, indicating that 
there could be 4--5 normal outbursts between two superoutbursts,
i.e., $N_{\rm nor}$ = 4--5.

We conduct polynomial fits to each light curve to estimate
key light curve parameters of the outbursts in Figure~\ref{LCs} \citep[see][for example]{Lee19}, such as peak magnitude, epoch of peak brightness, etc.
As shown in the bottom panel of Figure~\ref{LCs} where each of the superoutbursts are aligned with respect to their respective peaks, the transition from the precursor to superoutburst occurs differently for each superoutburst, although they have very similar light curve shapes overall.
The estimated outburst amplitude, duration as well as rising
and decline rate of \target\ are comparable to those observed in
SU UMa and U Gem dwarf novae sampled in \citet{Otu16}.
\citet{Osa96} classified the three (``active'',``intermediate'',``WZ Sge'') activity subgroups of SU UMa-type dwarf novae based on $P_{sc}$ and $N_{\rm nor}$ which are related with mass-transfer rate. According to this classification, $P_{sc}$ and $N_{\rm nor}$ of {\target} are similar to those of ``active'' SU UMa-type dwarf novae with a large mass-transfer rate.

SU UMa-type dwarf novae show superhumps, 
a periodic variation of brightness during superoutbursts,
whose presence is often investigated using power spectrum analysis 
of the plateau phase \citep{Kat09}.
After the best-fit polynomial to each superoutburst light curve
is subtracted out from the light curve, we combined residual light curves of the plateau phases 
(double arrow in the bottom panel of Figure~\ref{LCs}) of the three superoutbursts in $BVI$.
since the superhump periods of the plateau phases are known to be 
relatively invariant among different superoutbursts of the same object \citep{Kat09}.
With the combined residual light curve, we attempt to measure the superhump period ($P_{\rm sh}$) of {\target}
using Lomb-Scargle analysis\footnote{We use astropy.stats.LombScargle for this. \citep[\url{http://docs.astropy.org/en/stable/stats/lombscargle.html};][]{Ast13,Ast18}}.

\begin{figure}
\epsscale{1.1}
\plotone{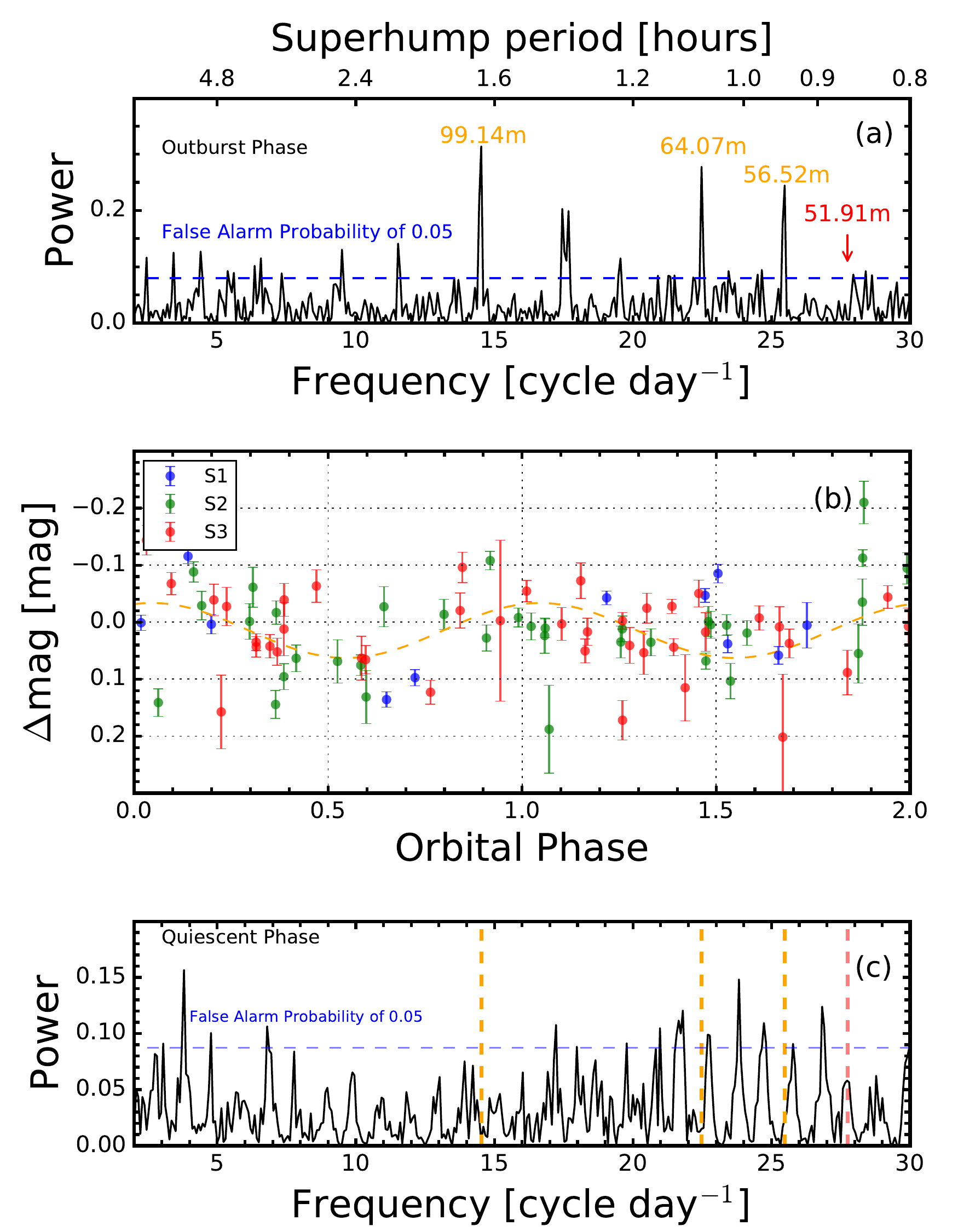}
\caption{(a) Power spectrum of superhump of \target\ from 
Lomb-Scargle Periodogram analysis of the combined light curve
of the three superoutbursts in $BVI$.
The horizontal dashed line is a false alarm probability of $0.05$.
Red arrow indicates the orbital period measured from the spectroscopic data. 
(b) Orbital phase diagram of the $V$-band residuals obtained after subtracting the 
polynomial fit from the observed light curves in Figure~\ref{LCs} (bottom panel).
We used the orbital period of 51.91 minutes.
(c) Power spectrum in quiescent phase of {\target} with the combined light curves 
between MJD of 57800 days and 57850 days in $BVI$.
The horizontal dashed line is a false alarm probability of $0.05$.
Four vertical dashed lines from left to right are 14.525, 22.475, 25.475 and 27.740 cycle day$^{-1}$.
}
\label{freq}
\end{figure}

Figure~\ref{freq}(a) shows the power spectrum of \target\
with three major powers at $14.525 \pm 0.075$, $22.475 \pm 0.075$, and $25.475 \pm 0.075$ cycle day$^{-1}$,
corresponding to $99.14 \pm 0.19$, $64.07 \pm 0.19$, and $56.52 \pm 0.19$ minutes, respectively.
We note that all these three potential periods of \target\ 
are below the period gap (2h--3h) of dwarf novae \citep{Otu16},
consistent with the postulation that \target\ is an SU UMa-type dwarf nova \citep{Kat09}.

We also calculated the power spectrum with $BVI$ light curves subtracted quiescent magnitudes in the quiescent phase 
between MJD of 57800 days and MJD of 57850 days.
The results are shown in Figure~\ref{freq}(c). 
The significant peaks are populated around a frequency of 25 cycle day$^{-1}$.
This seems to imply that an orbital period of {\target} is around 58 minutes and the superhump period will be the similar value.
Since, in Section~\ref{sec:spec}, a spectroscopic orbital period is $51.91 \pm 2.50$ minutes,
the closest superhump period of $56.52 \pm 0.19$ minutes from the spectroscopic orbital period is the most possible superhump period.
Orbital phase diagram of the $V$-band for the orbital period of $51.91$ minutes are shown in Figure~\ref{freq}(b).

\section{Spectroscopic analysis: Helium to Hydrogen Ratio and Orbital Period}\label{sec:spec}

We conduct time series spectroscopic observations of \target\
using GMOS on the 8-m Gemini South telescope on 2018 February 9th 
around the end of the $S2$ superoutburst as shown in Figure~\ref{LCs}.
We obtain 4 blue (3725{\AA}--6770{\AA}) and 4 red (5260{\AA}--9860{\AA}) GMOS spectra, each with an exposure time of 400 seconds, during about 1 hour.
The spectral resolutions are ${\rm FWHM} = 2.73 {\rm \AA}$ at $4610 {\rm \AA}$ in the blue spectra and ${\rm FWHM} = 3.98 {\rm \AA}$ at $7640 {\rm \AA}$ in the red spectra.
In addition, we make observations of the target using LDSS3-C on the 6.5-m Magellan Clay
telescope on 2017 April 30th around the end of the $N2$ normal outburst  
and we take two 1200-s exposures of LDSS3-C covering the spectral range of 4400{\AA}--9990{\AA}.
The spectral resolution is ${\rm FWHM} = 8.25 {\rm \AA}$ at $7100 {\rm \AA}$.

\begin{figure}
\epsscale{1.2}
\plotone{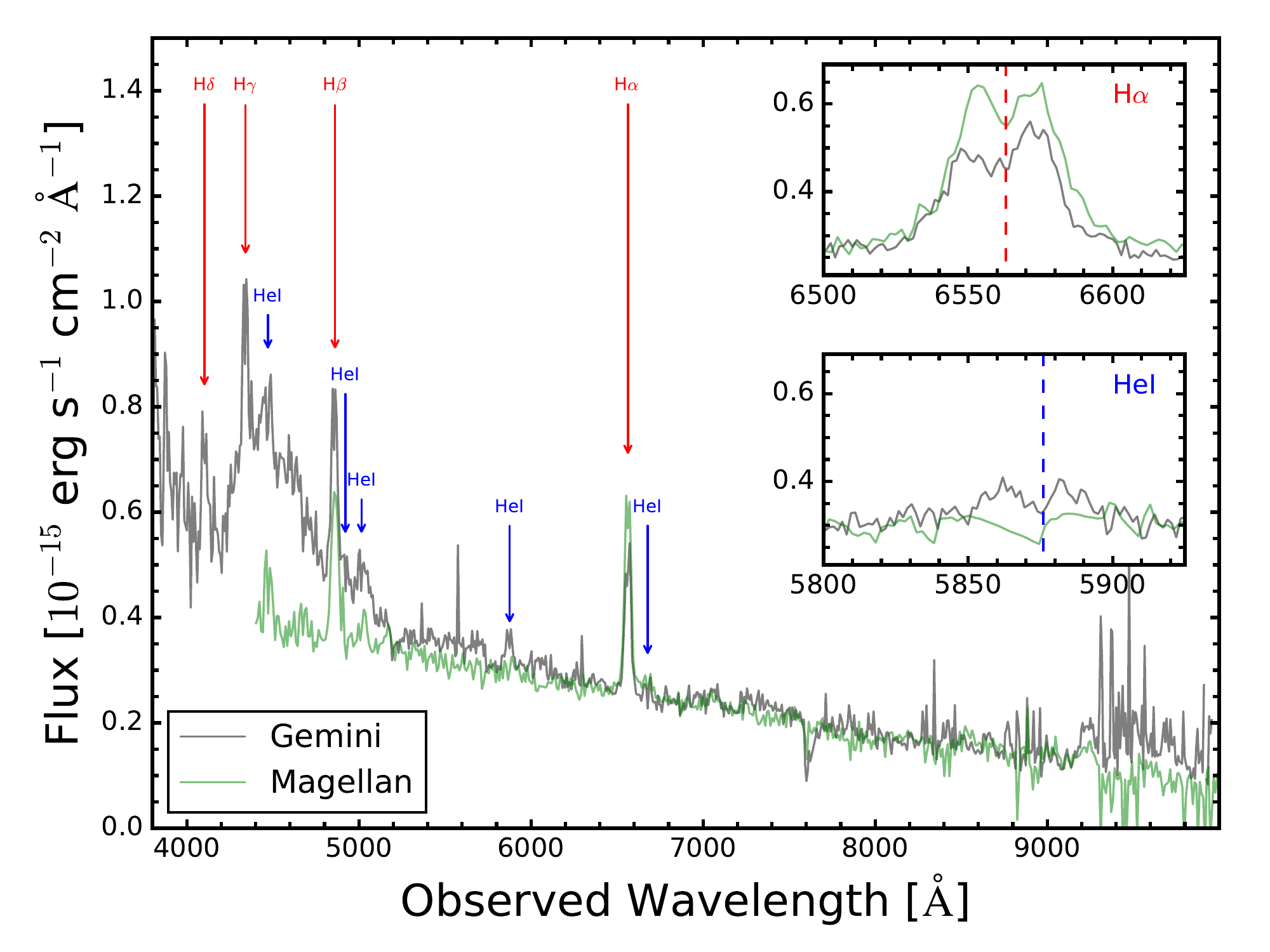}
\caption{Stacked Gemini (gray) and Magellan (green) spectra of \target.
The identified Hydrogen and Helium emission lines are denoted by the arrows.
The inset panels give a magnified view of the H$\alpha$ and the {\hei} line, respectively, confirming its  
double-peaked structure. 
The Magellan spectrum was obtained when the source was fainter than the epoch
of Gemini observations, and is shifted upward to match the continuum level 
of the Gemini spectrum.}
\label{spec}
\end{figure}

Figure~\ref{spec} shows our stacked spectra of \target: 
gray color for the Gemini stacked spectrum and green color for the Magellan one.
The overall continua of these spectra are almost same, but Gemini spectrum shows a flux excess below wavelength of 5500 ${\rm \AA}$. 
This seems to imply that Gemini spectrum is weakly affected by the outburst, while Magellan data is a quiescent spectrum.
We identify strong Balmer emission lines
alongside weak {\hei} lines.
The insets in Figure~\ref{spec} are an enlarged spectrum around the H$\alpha$ and ${\hei}_{\lambda5876}$ lines,
clearly revealing a double-peak structure of the line 
which is indicative of the presence of rotating accretion \citep{Sha83}.
We estimate the He to H line ratio (He/H) of \target\ using 
the ${\hei}_{\lambda5876}$ and the H$\alpha$ lines which have been
used in \citet{Ken15}.
The flux of the two emission lines are obtained after 
the underlying continuum emission is subtracted
by a polynomial fit obtained from the line-free
parts of the spectrum. 
The mean estimated He/H ratios in the eight 
individual GMOS spectra are $0.18$ $\pm$ $0.08$ in the range of 0.07--0.29.

We use the radial velocity variations of the H$\alpha$ emission line
in the eight GMOS time series spectra to estimate 
the orbital period of the primary of \target\
by applying the antisymmetric two Gaussian convolution technique \citep{Sha83} 
to the observed spectra in the range of 6500{\AA}--6600{\AA}. 
We fit the observed spectra of the H$\alpha$ lines in the eight Gemini spectra
with 13{\AA}--21{\AA} for the range of 
the half separation distance between the two Gaussian peaks,
10{\AA}--50{\AA} for the range of the FWHM of the two Gaussians,
and 6500{\AA}--6600{\AA} for the central wavelength range of the H$\alpha$ line.
We, then, adopt the circular orbit as following function

\begin{equation}\label{eq1}
v(t) = \gamma + K\sin[2\pi(t-T_0)/P_{\rm orb}]
\end{equation}

\noindent
where $\gamma$, $K$, $T_0$, and $P_{\rm orb}$ 
are the systematic velocity, 
semi-amplitude of radial velocity of the primary, 
time of the conjunction, 
and orbital period, respectively,
for our fitting of the measured radial velocities of the H$\alpha$ lines.

We followed the method in \citet{Aug96} that determines the best-fit
$K$ and $P_{\rm orb}$ when the uncertainty to magnitude
ratio (= $\sigma_{K}$/$K$) of the parameter $K$ is the minimum.
We obtain the minimum $\sigma_K/K$ 
when the separation of the two Gaussians in the H$\alpha$ line is 15.5{\AA},
and this gives the best-fit parameters 
$\gamma = 38.5 \pm 12.4$ km s$^{-1}$, $K = 132.9 \pm 18.2$ km s$^{-1}$, and $P_{\rm orb} = 51.91 \pm 2.50$ minutes.

\begin{figure}
\epsscale{1.2}
\plotone{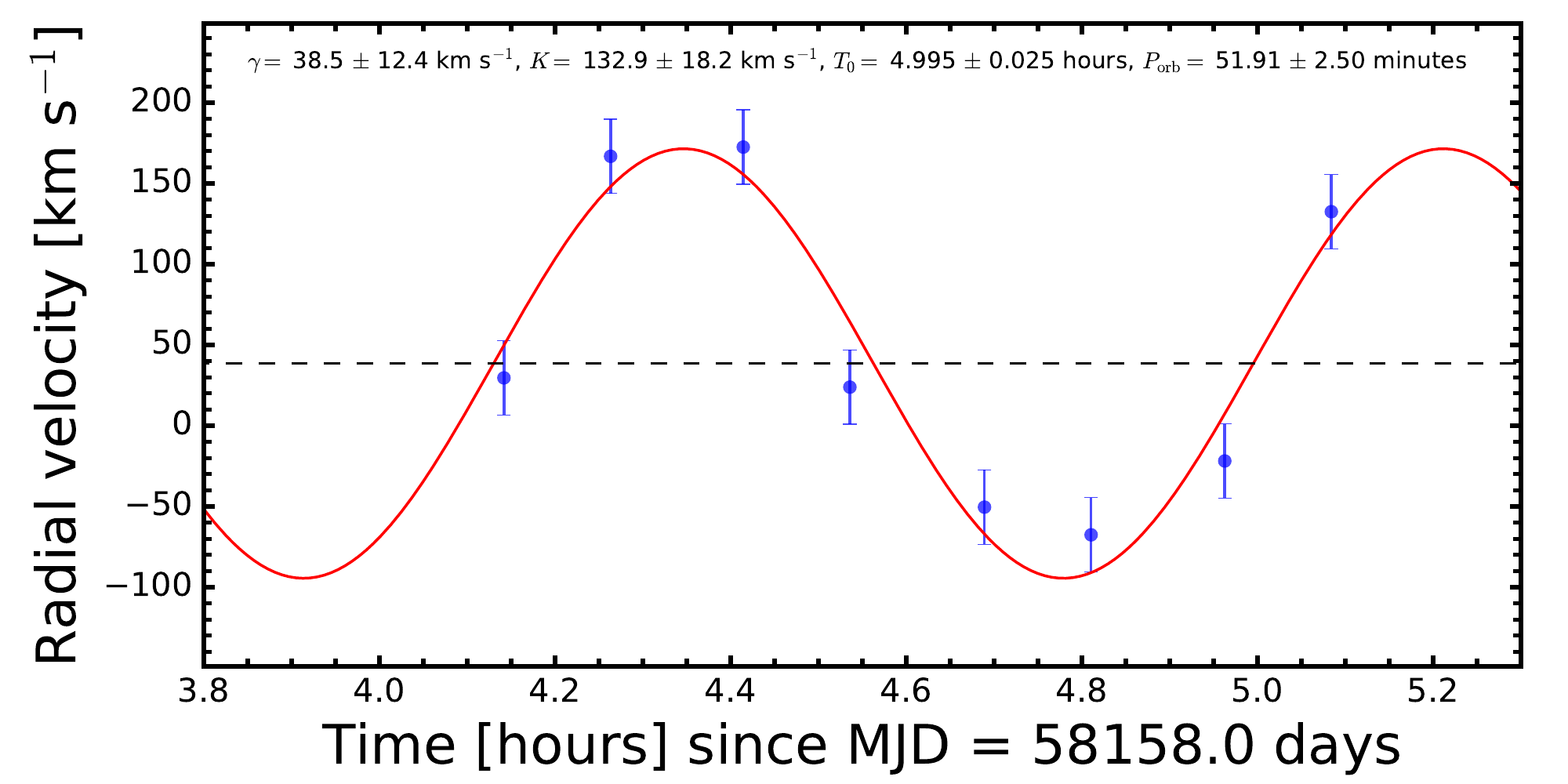}
\caption{Observed radial velocity (blue circles) sequence of H$\alpha$ line from the Gemini spectra.
The red curve represents the best-fit orbital velocity.
The dashed line is the systematic velocity of the binary.
Uncertainties of radial velocity at each epoch are the interval (0.5${\rm \AA}$) of the half separation distance ($a$) which is converted to $23$ km s$^{-1}$ at $6563{\rm \AA}$.
}
\label{scurve}
\end{figure}

Figure~\ref{scurve} shows that the fitted circular orbit (red curve) of Equation~(\ref{eq1})
matches well the observed radial velocities (blue circles) of \target.
In the figure, the first four spectra show the
mean He/H ratio of $\sim$ 0.12,
while that of the next three is 0.26.
We attribute the apparent systematic difference of the mean 
He/H ratio to 
the interference by the NaD ($5890$${\rm \AA}$ and $5896$${\rm \AA}$) 
from the cool secondary star to the broad ${\hei}_{\lambda5876}$ line from the accretion disk 
when the secondary star is moving toward us
in the first four spectra.
The systematic difference in the line ratios between the first four
and the next three spectra also supports the validity of our estimation
of the orbital period of 51.91 minutes.
Finally, we will adopt 0.26$\pm$0.04 as the uncontaminated He/H of {\target}.
It should be noticed that He/H ratio measured from Gemini spectra could be affected by the superoutburst since we observed just before the end of the superoutburst (see Figure~\ref{LCs} and \ref{spec}).
It may make additional uncertainty about the measurement of He/H. 
Although it is difficult to measure this uncertainty, the measured He/H is probably weakly affected by outburst since a flux excess is weak above wavelength of $5500{\rm \AA}$ comparing a quiescent spectrum from the Magellan observations. In the case of a quiescent spectrum from Magellan,  He/H ratio ($0.10\pm0.01$) is quiet low.
It seems that He emission line  is partly contaminated by the NaD absorption since the exposures are three times longer (20 minutes) than those of the Gemini observations.

\section{Discussion} \label{sec:discussion}

\begin{figure*}
\epsscale{1.2}
\plotone{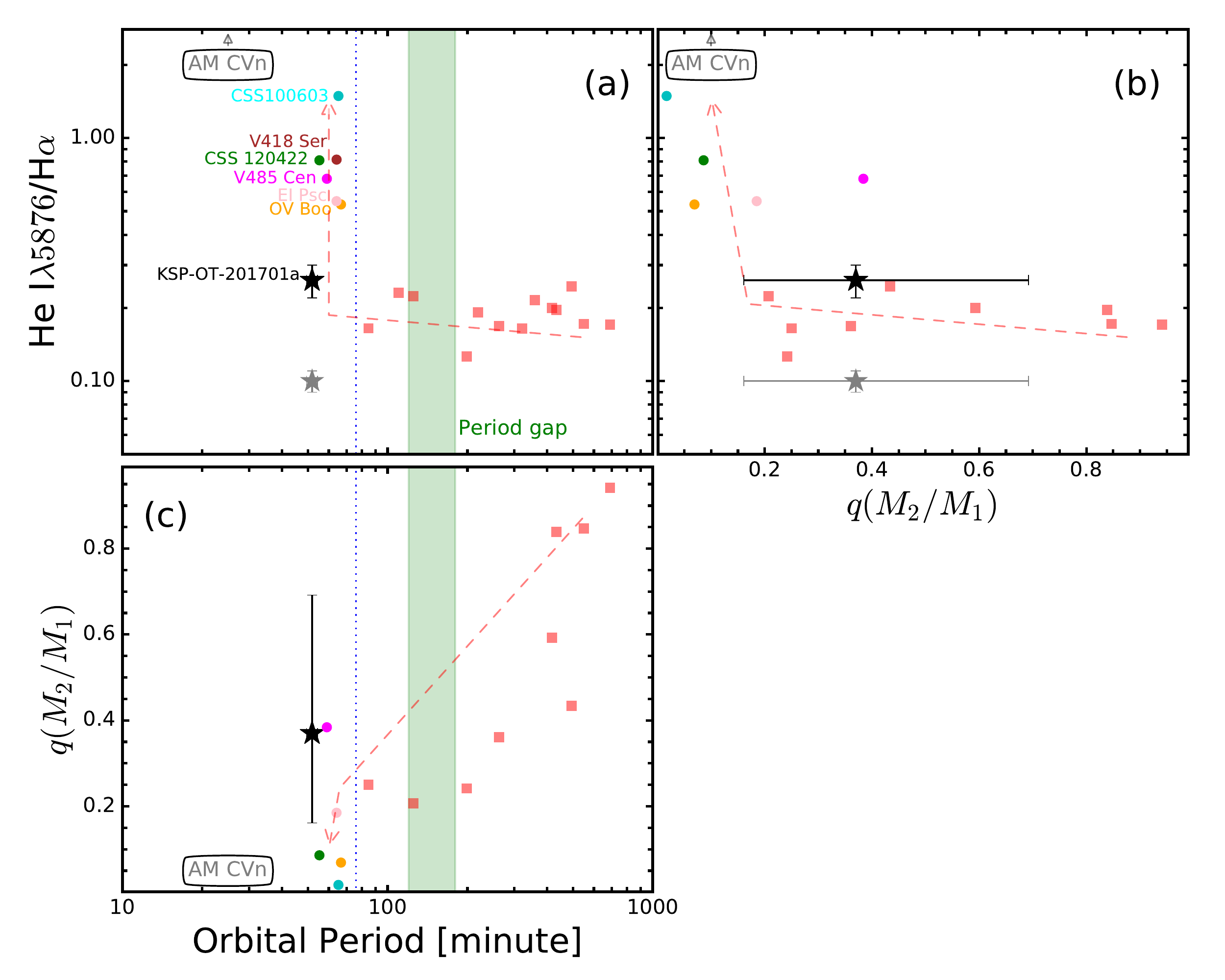}
\caption{Comparisons of the orbital periods, He/H,
and mass ratio $q(M2/M1)$ of short-period dwarf novae (colored filled circles), 
long-period dwarf novae (red squares), and
 \target\ at an outburst (black star) and a quiescent phase (gray star) :
(a) Orbital periods vs. He/H ratios;
(b) mass ratios vs. He/H ratios;
(c) orbital periods vs. mass ratios.
Mass ratio is a mean value measured by the three emperical relations.
The green shared area and the blue dotted vertical lines
in (a) and (c) are the period gap between 2 and 3 hours 
and the period minimum of 76 minutes, respectively. 
The red dashed lines with an arrow in (a)--(c) 
roughly represent the expected evolutionary paths
of a dwarf nova with an evolved secondary.
as its orbital period decreases as a result of mass loss.
The locations of AM CVn, 
which are a group of dwarf novae with small orbital periods,
small mass ratio, and no H emission, 
are shown in (a)--(c). 
See Section~\ref{sec:discussion} for the references of the dwarf novae
shown here. 
}
\label{prop}
\end{figure*}

In Figure~\ref{prop}, we compare the orbital period, 
He/H ratio, and mass ratio ($q=M_2/M_1$) of \target\ (filled star)
to those of 
He CVs (filled colored circles) of V485 Cen \citep{Aug96,Ole97}, EI Psc \citep{Tho02a}, OV Boo \citep{Pat08}, CSS 100603 \citep{Bre12,Kat10}, CSS 120422 \citep{Car13}, and V418 Ser \citep{Gre20,Ken15} and long-period dwarf novae \citep[red squares; ][]{Wil82,Rit03, Tho04}.
They represent the entire sample of the He CVs and long-period dwarf novae
observed with these parameters available in literature to the best of our knowledge.
Among the short-period dwarf novae,
\target\ in Figure~\ref{prop}(a) appears to have the lowest He/H ratio of 0.10--0.26 among 
the short-period dwarf novae (i.e., He CVs and AM CVn)
and it is similar to the ratios typically observed
in long-period dwarf novae (see Section~\ref{sec:spec}). 
According to binary population synthesis models \citep{Pod03,Gol15}, 
a long-period dwarf nova with an evolved secondary 
and a low central hydrogen fraction $X_c \lesssim 0.1$
evolves to He CVs as its orbital period and mass ratio decrease.
This indicates that \target\ is an intermediate stage 
between long-period dwarf novae and He CVs.

The superhump period excess, i.e., $\epsilon = P_{sh}/P_{orb} - 1$, 
can be derived from the precession of the eccentric disk of a dwarf nova 
caused by the secondary star perturbation, dependent on the mass ratio of the system
\citep{Whi88,Lub91,Hir90,Pat01,Pat05,Kni06}.
We follow three empirical relations 
between the period excess and mass ratio of dwarf novae
\citep{Pat01,Pat05,Kni06} 
to estimate the mass ratio of \target\ to be 
$0.32^{+0.46}_{-0.17}$, $0.37^{+0.23}_{-0.21}$, and $0.41^{+0.27}_{-0.24}$.
As shown in Figure~\ref{prop}(b),
the mass ratios of long-period dwarf novae are in the range of
 0.207--0.941 with an average of 0.523,
whereas those of He CVs are in the range of 0.017--0.384
with an average of 0.148.
The mass ratio of \target\ is comparable with what is typically
found in long-period dwarf novae, but it is very similar to
the largest ratio found in He CVs.

The dashed red lines in Figure~\ref{prop} show the potential evolutionary track of \target\
from a long-period dwarf nova with an evolved secondary to a He CV as observed today \citep{Pod03,Ken15}.
During the previous evolution to the current He CV phase,
the mass ratio and orbital period of \target\ have decreased
due to the mass loss of the secondary 
and the angular momentum loss of the binary system
to reach the current orbital period of $\sim$ 52 minutes
below the period minimum.
It is highly likely that \target\ is in an early phase of a He CV given 
its small He/H ratio and large mass ratio, 
having lost only part of its hydrogen envelop so far.
We expect \target\ to continuously lose its hydrogen envelope,
 progressively revealing its helium envelope with an increasing He/H ratio, 
as commonly observed in He CVs before they become AM CVn.
The expected sudden increase of He/H ratio deep inside a
low mass star as a result of nucleosynthesis \citep[see][]{Boe92}
leads to the large He/H ratio observed in He CVs as the
secondaries keep losing their hydrogen envelopes \citep{Sch02}.

\section{Summary and Conclusion} \label{sec:summary}

We discovered for the first time a short-period and unusually 
helium-deficient dwarf nova ({\target}).
We summarize and conclude our results as follows.

\begin{itemize}
\item \target\ has shown rich features of outbursts of a dwarf nova, including
normal outbursts and superoutbursts with precursors. 
In terms of superoutburst cycle ($P_{sc}$) and the number of normal outbursts ($N_{\rm nor}$), {\target} has similar features of the "active" subgroup of SU UMa dwarf novae with large mass-transfer rate.

\item We determine the orbital period of \target\ to be 51.91 minutes based on
our time-series spectroscopic observations and the period of superhumps 
to be $56.52 \pm 0.19$ minutes during the plateau of superoutbursts.
The superhump period excess is large ($0.089 \pm 0.053$), 
compatible with a mass ratio ($M_2/M_1$) of $0.37^{+0.32}_{-0.21}$.
\item  The observed flux ratios (${\hei}_{\lambda5876}/{\ha} = 0.10\pm0.01$ at a quiescent phase and $0.26\pm0.04$ at an outburst phase)
of \target\ are similar to those of dwarf novae with a long ($>$ 80 minutes) period, and is the lowest value among He CVs.
\item We suggest that the short-period and unusually helium-deficient nature of \target\ 
is consistent with a dwarf nova in an intermediate stage 
evolving into a He CV from a long-period dwarf nova 
with an evolved secondary star.
\end{itemize}

\begin{acknowledgments}
This research has made use of the KMTNet facility operated by the Korea Astronomy and Space Science Institute, and the data were obtained at the three host sites of CTIO in Chile, SAAO in South Africa, and SSO in Australia.
YDL acknowledges support from Basic Science Research Program through the National Research Foundation of Korea (NRF) funded by the Ministry of Education (2020R1A6A3A01099777). 
DSM was supported in part by a Leading Edge Fund from the Canadian Foundation 
for Innovation (project No. 30951) and a Discovery
Grant (RGPIN-2019-06524) from the Natural Sciences and Engineering Research Council (NSERC) of Canada. 
HSP was supported in part by the National Research Foundation of Korea (NRF) grant funded by the Korea government (MSIT, Ministry of Science and ICT; No. NRF-2019R1F1A1058228).
MRD acknowledges support from NSERC through grant RGPIN-2019-06186, 
the Canada Research Chairs Program and  
the Canadian Institute for Advanced Research. 
\end{acknowledgments}

\bibliography{draft}

\end{document}